\documentclass[12pt,tightenlines]{iopart}
\usepackage{color}
\usepackage{graphicx}
\usepackage{epsf}
\usepackage{graphicx,epsfig}
\usepackage{iopams}
\usepackage{setstack}
\begin{document}

\title{Second-order matter perturbations in a $\Lambda$CDM cosmology and non-Gaussianity}

\vspace{0.8cm}

\author{ Nicola Bartolo$^{1,2}$, Sabino Matarrese$^{1,2}$, Ornella Pantano$^{1}$ and 
Antonio Riotto$^{2,3}$} 
\vspace{0.4cm}
\address{$^1$ Dipartimento di Fisica ``G. Galilei'', Universit\`{a} degli Studi di 
Padova, \\ via Marzolo 8, I-35131 Padova, Italy} 
\address{$^2$ INFN, Sezione di Padova, via Marzolo 8, I-35131 Padova, Italy}
\address{$^3$ CERN, Theory Division, CH-1211 Geneva 23, Switzerland}
\eads{\mailto{nicola.bartolo@pd.infn.it},~\mailto{sabino.matarrese@pd.infn.it},\\
\mailto{ornella.pantano@pd.infn.it} and \mailto{riotto@mail.cern.ch}}

\date{\today}
\vspace{1cm}

\begin{abstract}
We obtain exact expressions for the effect of primordial non-Gaussianity on the matter density perturbation 
up to second order in a $\Lambda$CDM cosmology, fully accounting for the general relativistic 
corrections arising on scales comparable with the Hubble radius. We present our results both in the 
Poisson gauge and in the comoving and synchronous gauge, which are relevant for comparison to different 
cosmological observables.
\\
\\
\\
CERN-PH-TH/2010-044
\end{abstract}

\maketitle

\section{Introduction} 

In this paper we discuss how primordial non-Gaussianity (NG) in the 
cosmological perturbations is left imprinted in the Large-Scale Structure (LSS) of the universe in a $\Lambda$CDM cosmology.
We show how the information on the
primordial non-Gaussianity, set on super-Hubble scales, flows into smaller scales 
through a complete General Relativistic (GR) computation. 
Primordial NG thus leaves an observable imprint in the LSS. 
Another interesting finding is that, on sufficiently large scales, there is another additional source of non-Gaussianity 
which arises from GR corrections, the leading contributions of which are Post-Newtonian terms as first pointed out 
in Ref.~\cite{noiLSS}.   
The importance of the signatures of primordial non-Gaussianity in the evolution of the matter density perturbations is due to the fact that 
future high-precision measurements of the statistics of  the dark matter density will allow to pin down the primordial non-Gaussianity,
thus representing a tool complementary to studies of the Cosmic Microwave Background (CMB) anisotropies. 
It is beyond the scope of this paper to go into the details of the various theoretical and observational methods 
related to this issue. This paper will then serve as a basic guideline to capture the starting point expressions that 
relates primordial NG to the dark matter density and gravitational potentials, outlining some specific non-Gaussian signatures in the LSS
that can be potentially interesting. We perform our calculations in assuming a flat Universe with pressure-less matter, 
i.e. Cold Dark Matter (CDM) plus non-relativistic 
ordinary matter and a cosmological constant, hereafter $\Lambda$CDM cosmology. 
We present our results both in the Poisson gauge and 
in the comoving time-orthogonal gauge, which are relevant for comparison to observations.

The primordial NG considered in our analysis is set at primordial (inflationary) epochs on large (super-Hubble) scales. At later times 
cosmological perturbations reenter the Hubble radius 
during the radiation or during the matter- and dark energy-dominated epochs. 
For scales re-entering the Hubble radius during 
the radiation dominated era one should include the radiation in the evolution 
equations, thus using a complete \emph{second-order} matter transfer function also for those scales. A detailed treatment of it 
has been given in Refs.\cite{CMB1,CMB2}. Refs.~\cite{pitrou,Fitz} also investigate the evolution of the dark matter 
perturbations up to second-order accounting for a radiation-dominated epoch, both analytically and numerically. 
Here we will focus on large scales for which the effects arising during the radiation-dominated epoch can be neglected. 

The plan of the paper is as follows. In Section 2 we give the general form of the perturbed line-element and we introduce  
the gauge-invariant curvature perturbation of uniform density hypersurface 
at first and second order, which is used to provide the inflationary initial 
conditions, including the effect of primordial non-Gaussianity. 
In Section 3 we derive the second-order expression for the density perturbation in a $\Lambda$CDM cosmology, 
in the Poisson gauge, taking into full account 
both NG initial conditions  and Post-Newtonian corrections arising from the non-linear evolution of perturbations 
according to the fully General Relativistic treatment. 
Section 4 contains a similar calculation in the comoving-synchronous gauge. Section 5 contains our concluding remarks. 

\section{Metric perturbations and primordial non-Gaussianity}
We consider a spatially flat Universe filled with a cosmological constant 
$\Lambda$ and a non-relativistic pressureless fluid of Cold Dark Matter (CDM), whose energy-momentum tensor reads 
$T^{\mu}_{~\nu}=\rho u^{\mu} u_{\nu}$. Following the notations of Ref.~\cite{MMB}, the perturbed line element around 
a spatially flat FRW background reads
\begin{equation}
\label{metric}
ds^2=a^2(\tau)\{-(1+2\phi)d\tau^2 + 2 \hat{\omega}_i d\tau dx^i+[(1-2\psi) \delta_{ij} + \hat{\chi}_{ij}]dx^i dx^j \}\, .
\end{equation}
where $a(\tau)$ is the scale factor as a function of conformal time $\tau$. 
Here each perturbation quantity can be expanded into a 
first-order (linear) part and a second-order contribution, as for example, 
the gravitational potential $\phi=\phi^{(1)}+\phi^{(2)}/2$. 
Up to now we have not choosen any particular gauge. 
We can employ the standard split of the perturbations into the so-called scalar, 
vector and tensor parts,  
according to their transformation properties with respect to 
the $3$-dimensional space with metric $\delta_{ij}$, where scalar parts are 
related to a scalar potential, vector parts to transverse (divergence-free) 
vectors and tensor parts to transverse trace-free tensors. 
Thus $\phi$ and $\psi$, the gravitational potentials, are scalar perturbations, 
and for instance, $\hat{\omega}_i^{(r)}=\partial_i\omega^{(r)}+\omega_i^{(r)}$, 
where $\omega^{(r)}$ is the scalar part and $\omega^{(r)}_i$ 
is a transverse vector, {\it i.e.} $\partial^i\omega^{(r)}_i=0$ ($(r)=(1,2)$ stand for the 
$r$th-order of the perturbations). The symmetric traceless tensor $\hat{\chi}_{ij}$ generally contains a 
scalar, a vector and a tensor contribution, namely $\hat{\chi}_{ij} = D_{ij} \chi + 
\partial_i \chi_j + \partial_j \chi_i + \chi_{ij}$, where $D_{ij} \equiv \partial_i \partial_j - (1/3) \nabla^2 \delta_{ij}$,
$\chi_i$ is a solenoidal vector ($\partial^i \chi_i=0$) and  $\chi_{ij}$ represents a traceless and 
transverse (i.e. $\partial^i \chi_{ij} =0$) tensor mode\footnote{In what follows, for our purposes 
we will neglect linear vector modes since they are not produced in standard 
mechanisms for the generation of cosmological perturbations (as inflation), 
and we also neglect tensor modes at linear order, since they give a negligible 
contribution to LSS formation.}. 
As for the matter component we split the mass density into a homogeneous 
$\rho(\tau)$ and a perturbed part as 
$\rho({\bf x},\tau)=\rho(\tau) (1+\delta^{(1)}+\delta^{(2)}/2)$ 
and we write the four velocity as $u^{\mu}=(\delta^{\mu}_0+v^{\mu})/a$ with $u^{\mu}u_{\mu}=-1$ and 
$v^\mu=v^{(1)\mu}+v^{(2)\mu}/2$.

The Friedmann background equations are 
\begin{equation}
3{\mathcal H}^2 =a^2(8 \pi G \rho(\tau)+\Lambda)\, , 
\end{equation}
and
\begin{equation} 
\rho'(\tau)=-3{\mathcal H} \rho(\tau) \, ,
\end{equation} 
where a prime stands for differentiation with 
respect to conformal time, and ${\mathcal H}=a'/a$. The matter density parameter is $\Omega_m(\tau)=8 \pi G a^2(\tau) \rho(\tau)
/(3 {\cal H}^2(\tau))$.

Before recalling how one can parametrize the primordial non-Gaussianity, we need to provide
a general definition for the amplitude of non-Gaussianity characterizing the matter density contrast beyond the usual 
second-order Newtonian contributions. It proves convenient to introduce an effective gravitational potential obeying the 
Poisson equation 
\begin{equation}
\label{P}
-\nabla^2 \Phi=\frac{3}{2} \Omega_m {\cal H}^2 \delta, 
\end{equation}
and at an initial epoch, deep in matter domination, we write  
\begin{equation}
\label{fnlphiin}
\Phi_{\rm in}=\Phi^{(1)}_{\rm in}+f_{\rm NL}(\Phi^{(1)2}_{\rm in}-\langle \Phi^{(1)2}_{\rm in} \rangle)\; ,
\end{equation}
with the dimensionless non-linearity parameter $f_{\rm NL}$ setting the level of quadratic non-Gaussianity, and $\Phi \propto g(\tau)$, $g(\tau)$ being the 
usual growth suppression factor (see Sec.~\ref{Poisson}). Therefore we will write 
the matter density contrast in Fourier space in 
terms of the linear density contrast by 
defining the kernel ${\cal K}_\delta({\bf k}_1,{\bf k}_2;\tau)$ depending on the 
wavevector of the perturbation modes as
\begin{eqnarray}
\label{defkerneldensity}
\delta_{{\bf k}} (\tau)&=& \delta^{(1)}_{{\bf k}} (\tau)+\frac{1}{2} \delta^{(2)}_{{\bf k}}(\tau)
\\
&= & \delta^{(1)}_{{\bf k}} (\tau)+\int \frac{d^3{\bf k}_1d^3{\bf k}_2}{(2 \pi)^3} \,{\cal K}_{\delta}({\bf k}_1,{\bf k}_2;\tau) \delta^{(1)}_{{\bf k}_1}(\tau) \delta^{(1)}_{{\bf k}_2}(\tau)
\delta_D ({\bf k}_{1}+{\bf k}_{2}-{\bf k})\, .\nonumber
\end{eqnarray}
We can write the kernel as 
\begin{eqnarray}
\label{kddthth}
{\cal K}_{\delta}({\bf k}_1,{\bf k}_2;\tau)  &=&  {\cal K}_{\delta}^N({\bf k}_1,{\bf k}_2;\tau)+\frac{3}{2} \Omega_m {\cal H}^2 f_{\rm NL}({\bf k}_1,{\bf k}_2,\tau) \frac{g_{\rm in}}{g(\tau)} 
\frac{k^2}{k_1^2 k_2^2}\, ,
\end{eqnarray}
where $k^2 \equiv |{\bf k}_1 + {\bf k}_2|^2$ and ${\cal K}_{\delta}^N({\bf k}_1,{\bf k}_2;\tau)$ is the second-order Newtonian kernel.

\subsection{Primordial non-Gaussianity and initial conditions}
We conveniently fix the initial conditions at the time when the cosmological perturbations relevant for LSS are outside the horizon.  
A standard and convenient way to account for any initial primordial non-Gaussianity is to consider the curvature perturbation 
of uniform density hypersurfaces $\zeta=\zeta^{(1)}+\zeta^{(2)}/2+\cdots$, where $\zeta^{(1)}=-\hat{\psi}^{(1)}-{\mathcal H} {\delta \rho}^{(1)}/{\rho}'$ and at second-order
~\cite{mw,review}
\begin{eqnarray}
\label{zeta2}
-\zeta^{(2)}&=& 
\hat{\psi}^{(2)}+{\mathcal H}\frac{\delta^{(2)}\rho}{\rho^\prime}
-2{\mathcal H}\frac{\delta^{(1)}\rho^\prime}{\rho^\prime}
\frac{\delta^{(1)}\rho}{\rho^\prime} 
-2\frac{\delta^{(1)}\rho}{\rho^\prime}({\hat \psi}^{(1)\prime}
+2 {\mathcal H} {\hat \psi}^{(1)})
\nonumber \\
&+&\left(\frac{\delta^{(1)} \rho}{\rho'}\right)^2 
\left({\mathcal H} \frac{\rho^{\prime\prime}}{\rho^\prime}-
{\mathcal H}^\prime-2{\mathcal H}^2\right) \, ,
\end{eqnarray}
where $\hat{\psi}^{(r)}=\psi^{(r)}+\nabla^2 \chi^{(r)}/6$. This is a gauge-invariant quantity which remains constant on super-horizon 
scales after it has been generated during a primordial epoch (and possible isocurvature perturbations are no longer present). 
Therefore,  $\zeta^{(2)}$ provides all the necessary information about the primordial level of non-Gaussianity. 
The conserved value of the curvature perturbation $\zeta$
allows to  set the initial conditions for the metric and matter perturbations accounting for the 
primordial contributions.
Different scenarios for the generation are characterized by different values of  $\zeta^{(2)}$, while
the post-inflationary nonlinear evolution due to gravity is 
common to all of them~\cite{BMR2,BMR3,BMR4,review}.
For example, in standard single-field inflation $\zeta^{(2)}$ is generated during inflation
and its value is $\zeta^{(2)}=2\left( 
\zeta^{(1)} \right)^2+{\cal O}\left(\epsilon,\eta\right)$~\cite{ABMR,Maldacena,BMR2}, where $\epsilon$ and $\eta$ are the usual slow-roll parameters. 
Therefore it turns out to be convenient to parametrize the primordial non-Gaussianity level 
in terms of the conserved curvature perturbation as in Ref. \cite{prl}  
\begin{equation}
\label{param}
\zeta^{(2)}=2 a_{\rm NL}\left(\zeta^{(1)}\right)^2\, ,
\end{equation}
where the parameter $a_{\rm NL}$ depends on the physics of a given scenario, and in full generality it can 
depend on scale and configuration. 
For example in the standard scenario $a_{\rm NL}\simeq 1$, while in the  
curvaton case (see e.g. Ref.~\cite{review})
$a_{\rm NL}=(3/4r)-r/2$, where 
$r \approx (\rho_\sigma/\rho)_{\rm D}$ is the relative   
curvaton contribution to the total energy density at curvaton 
decay. In the minimal picture for the inhomogeneous 
reheating scenario, $a_{\rm NL}=1/4$. For other scenarios we refer 
the reader to Ref.~\cite{review,reviewT,reviewChen,reviewKoyama,reviewByrnes}. 
One of the best techniques 
to detect or constrain the primordial large-scale non-Gaussianity is 
through the analysis 
of the CMB anisotropies, for example by studying the CMB bispectrum
~\cite{review,S1,S2}. The non-linearity parameter $f_{\rm NL}$ as defined in Eq.~(\ref{fnlphiin}) is defined also 
to make contact with the primordial non-Gaussainity entering in the CMB 
anisotropies. For large primordial non-Gaussianity, when $|a_{\rm NL} | \gg 1$, 
$f_{\rm NL} \simeq 5 a_{\rm NL}/3$~(see Refs.~\cite{review,prl}). 
  
\section{Dark matter density perturbations at second-order: Poisson gauge}
\label{Poisson}

The goal of this section is to compute the matter density contrast in the Poisson gauge~\cite{Bert}, 
{\it i.e.} the generalization beyond linear order of the longitudinal gauge, 
by which a more direct comparison with the standard Newtonian 
approximation adopted in the interpretation of LSS observations and in N-body simulations in Eulerian coordinates s possible (see, however, 
Ref.~\cite{yoo} for a critical discussion of the potential problems connected to the use of the Poisson gauge on scales comparable 
with the Hubble radius).   
In the Poisson gauge one scalar degree of freedom is eliminated from the $g_{0i}$ component of the metric, 
and one scalar and two vector degrees of freedom are eliminated from $g_{ij}$.

Let us briefly recall the results for the linear perturbations in 
the case of a non-vanishing cosmological $\Lambda$ term. 
At linear order the traceless part of the ($i$-$j$)-components of
Einstein equations gives 
$\phi^{(1)}=\psi^{(1)} \equiv \varphi$. Its trace gives the evolution 
equation for the linear 
scalar potential $\varphi$
\begin{equation}
\label{ev}
\varphi''+3 {\mathcal H} \varphi'+a^2 \Lambda \varphi=0\, .
\end{equation}
Selecting only the growing mode solution one can write
\begin{equation}
\label{relphiphi_0}
\varphi({\bf x}, \tau)= g(\tau)\, \varphi_0({\bf x}) \, ,
\end{equation}
where $\varphi_0$ is the peculiar gravitational potential linearly 
extrapolated to the 
present time $(\tau_0)$ and $g(\tau)=D_+(\tau)/a(\tau)$ is the so called 
growth-suppression factor, where $D_+(\tau)$ is the usual linear growing-mode 
of density fluctuations in the Newtonian limit, i.e. the no-decaying solution of the differential equation
\begin{equation}
\label{D+}
D^{\prime\prime} + {\cal H} D^\prime - \frac{3}{2} {\cal H}^2 \Omega_m D =0 \;.
\end{equation}
We normalize the growth factor so that $D_+(\tau_0)=a_0=1$. The exact form of $g$ can be found in 
Refs.~\cite{lahav,Carroll,Eisenstein}. 
In the $\Lambda=0$ case $g=1$. A very good approximation for $g$ as a 
function of redshift $z$ is given in 
Refs.~\cite{lahav,Carroll} 
\begin{equation}
g \propto \Omega_m\left[\Omega_m^{4/7} - \Omega_\Lambda +
\left(1+ \Omega_m/2\right)\left(1+ \Omega_\Lambda/70\right)\right]^{-1} \;, 
\end{equation}
with $\Omega_m=\Omega_{0m}(1+z)^3/E^2(z)$, 
$\Omega_\Lambda=\Omega_{0\Lambda}/E^2(z)$, 
$E(z) \equiv (1+z) {\mathcal H}(z)/{\mathcal H}_0 = \left[\Omega_{0m}(1+z)^3 + 
\Omega_{0\Lambda}\right]^{1/2}$ and 
$\Omega_{0m}$, $\Omega_{0\Lambda}=1-\Omega_{0m}$, the present-day
density parameters of non-relativistic matter and cosmological constant, 
respectively. According to our normalization, 
$g(z=0)=1$. 
The energy and momentum constraints provide the density and
velocity fluctuations in terms of $\varphi$ 
(see, for example, Ref.~\cite{BMR2} and~\cite{mhm,tomita1} 
for the $\Lambda$ case)
\begin{eqnarray} 
\label{lindensvel}
\delta^{(1)}& = & \frac{1}{4 \pi G a^2 \bar\rho} 
\left[ \nabla^2 \varphi - 3{\cal H}\left(\varphi' + {\cal H} 
\varphi\right) \right], \\
\label{v1}
v^{(1)}_{i} & = & - \frac{1}{4 \pi G a^2 \bar\rho} \partial_i 
\left(\varphi' + {\cal H} \varphi\right) \;.  
\end{eqnarray}

In a similar way the expression of the second-order matter density contrast can be computed starting from the energy constraint given by the 
$(0-0)$ Einstein equation, once the evolution of the gravitational potentials is known. The latter has been already computed in 
detail in Ref.~\cite{Full} and in the following we summarize the main results. 

The evolution equation for the second-order gravitational potential $\psi^{(2)}$ 
is obtained from the trace of the ($i$-$j$)-Einstein equations \footnote{
The second-order perturbations of the Einstein tensor $G^\mu_{~\nu}$ 
can be found for any gauge in 
Appendix A of Refs.~\cite{ABMR,review} and directly in the Poisson gauge, e.g., in Refs.~\cite{CMB1,CMB2}. 
The perturbations of the 
energy-momentum tensor up to 
second order in the Poisson gauge have been computed in Ref.~\cite{BMR2} 
for a general perfect fluid 
(see Ref.~\cite{review} for expressions in any gauge).}    

\begin{equation}
\label{Psieq}
\psi^{(2)''}+3 {\mathcal H} \psi^{(2)'}+a^2 \Lambda \psi^{(2)}= S(\tau)\, ,
\end{equation}
where $S(\tau)$ is the source term   
\begin{eqnarray}
\label{PSIsource}
S(\tau)&=&g^2\Omega_m {\mathcal H}^2 \Bigg[\frac{(f-1)^2}{\Omega_m} 
\varphi_0^2+2 \Bigg(2 \frac{(f-1)^2}{\Omega_m} -\frac{3}{\Omega_m}+3 \Bigg) \alpha_0({\bf x}) 
\nonumber \\
&+& g^2 \Bigg[ \frac{4}{3}   
\left( \frac{f^2}{\Omega_m}+\frac{3}{2} \right) 
\nabla^{-2}\partial_i \partial^j
\left(\partial^i \varphi_0 \partial_j \varphi_0\right) - 
\left( \partial^i \varphi_0 
\partial_i \varphi_0 \right) \Bigg] \, ,
\end{eqnarray}
where, for simplicity of notation, we have introduced 
\begin{equation}
\label{alpha0}
\alpha_0({\bf x})=\Bigg[ \nabla^{-2} 
\left( \partial^i \varphi_0 \partial_i \varphi_0 \right)- 3 \nabla^{-4} \partial_i \partial^j
\left(\partial^i \varphi_0 \partial_j \varphi_0 \right)\Bigg]\, ,
\end{equation}
and
\begin{equation}
f(\Omega_m)=\frac{d \ln D_+}{d \ln a}=1+\frac{g'(\tau)}{{\mathcal H} g(\tau)}\, ,
\end{equation}
which can be written as a function of $\Omega_m$ as $f(\Omega_m) 
\approx \Omega_m(z)^{4/7}$
~\cite{lahav,Carroll}. In Eq.~(\ref{PSIsource}) $\nabla^{-2}$ stands 
for the inverse of the Laplacian operator. 

The solution of Eq.~(\ref{Psieq}) is then obtained using Green's method with growing and decaying solutions of the 
homogeneous equation $\psi_+(\tau)=g(\tau)$ and $\psi_-(\tau)={\mathcal H}(\tau)/a^2(\tau)$, respectively. The second-order 
gravitational potentials then read~\cite{Full}
\begin{eqnarray}
\label{PSI}
\fl \psi^{(2)}(\tau)&=&\left( B_1(\tau)-2g(\tau)g_{\rm in} -\frac{10}{3}(a_{\rm NL}-1)g(\tau)g_{\rm in} \right)\varphi_0^2 
+\left( B_2(\tau) -\frac{4}{3}g(\tau)g_{\rm in}  \right) \alpha_0({\bf x}) 
\nonumber\\
\fl &+& B_3(\tau) \nabla^{-2} \partial_i\partial^j(\partial^i \varphi_0 \partial_j 
\varphi_0 )+B_4(\tau) \partial^i \varphi_0 \partial _i\varphi_0 \, , 
\\
\fl
\label{PHI}
\phi^{(2)}(\tau)&=&\left( B_1(\tau)+4g^2(\tau)
-2g(\tau)g_{\rm in} -\frac{10}{3}(a_{\rm NL}-1)g(\tau)g_{\rm in}
\right)\varphi_0^2 
+\Bigg[ B_2(\tau)+\frac{4}{3} g^2(\tau)  
\nonumber \\
\fl &\times& \left( e(\tau)+\frac{3}{2} \right)
-\frac{4}{3}g(\tau)g_{\rm in} \Bigg] \alpha_0({\bf x}) 
+B_3(\tau) \nabla^{-2} \partial_i\partial^j(\partial^i \varphi_0 \partial_j 
\varphi_0 )+B_4(\tau) \partial^i \varphi_0 \partial _i\varphi_0\, , \nonumber \\
\fl
\end{eqnarray}
where we have introduced 
$B_i(\tau)={\mathcal H}_0^{-2} \left(f_0+3 \Omega_{0m}/2 \right)^{-1} 
\tilde{B}_i(\tau)$ with the following definitions
\begin{eqnarray}
\label{B1B2}
\fl \tilde{B}_1(\tau)&=&\int_{\tau_{\rm in}}^\tau d\tilde{\tau} \,{\mathcal H}^2(\tilde{\tau}) 
(f(\tilde{\tau})-1)^2 C(\tau,\tilde{\tau})\, , 
\\ \fl 
\tilde{B}_2(\tau)&=&2\int_{\tau_{\rm in}}^\tau d\tilde{\tau} \, {\mathcal H}^2(\tilde{\tau}) 
\Big[2 (f(\tilde{\tau})-1)^2-3
+3 \Omega_m(\tilde{\tau}) \Big] C(\tau,\tilde{\tau})\, , \\
\fl \tilde{B}_3(\tau)&=&\frac{4}{3} \int_{\tau_{\rm in}}^\tau d\tilde{\tau} \left(e(\tilde{\tau})
+\frac{3}{2} \right) C(\tau,\tilde{\tau}) \, , \,\,\,\,\,\,\,\,\,\,\,\,
\tilde{B}_4(\tau)= - \int_{\tau_{\rm in}}^\tau d\tilde{\tau} \,C(\tau,\tilde{\tau})\, ,
\end{eqnarray}
and 
\begin{equation}
C(\tau,\tilde{\tau})= g^2(\tilde{\tau}) a(\tilde{\tau}) 
\Big[ g(\tau){\mathcal H}(\tilde{\tau})-g(\tilde{\tau}) 
\frac{a^2(\tilde{\tau})}{a^2(\tau)} {\mathcal H}(\tau) \Big] \, ,
\end{equation}
with 
\begin{equation}
\label{e}
e(\Omega_m)  \equiv \frac{f^2(\Omega_m)}{\Omega_m} \, .
\end{equation}
The expression for $\phi^{(2)}$ is obtained from the 
relation between $\psi^{(2)}$ and $\phi^{(2)}$
\begin{equation}
\label{rela}
\fl \nabla^2 \nabla^2 \psi^{(2)}=\nabla^2 \nabla^2 \phi^{(2)} -4g^2 \nabla^2 \nabla^2 
\varphi^2_0 -\frac{4}{3} g^2 \Big( e + \frac{3}{2} \Big) 
\Big[ \nabla^2 (\partial_i \varphi_0 \partial^i \varphi_0) -3 \partial_i 
\partial^j (\partial^i\varphi_0 \partial_j \varphi_0) \Big]\, ,
\end{equation}
which follows from the traceless part of the ($i$-$j$)-component of 
Einstein equations~\cite{Full}. Here $\varphi_{\rm in}=g_{\rm in} \varphi_0$ represents the 
initial condition taken at some time $\tau_{\rm in}$ deep in the matter dominated era on super-horizon 
scales. 
The solutions~(\ref{PSI}) and~(\ref{PHI}) 
properly account for the non-Gaussian initial conditions parametrized by $(a_{\rm NL}-1)$. 
These are obtained using the expression for $\zeta^{(2)}$ during the 
matter-dominated epoch together with 
Eq.~(\ref{rela}) and 
the second-order ($0$-$0$)-component of Einstein equations 
(both evaluated for a matter-dominated epoch), so that one can express 
$\phi^{(2)}_{\rm in}$ and $\psi^{(2)}_{\rm in}$ in terms of $\zeta^{(2)}$ of Eq.~(\ref{param}), where $\zeta^{(1)}=-5 \varphi_{\rm in}/3$
(see Refs.~\cite{BMR2,BMR4,review,prl,Full}).

Before proceeding further notice that in the expression for the second-order gravitational potentials
of Eq.~(\ref{PSI}) and~(\ref{PHI}) we recognize two contributions. The term which  
dominates on small scales, $[B_3(\tau) \nabla^{-2} 
\partial_i\partial^j(\partial^i \varphi_0 \partial_j 
\varphi_0 )+B_4(\tau) \partial^i \varphi_0 \partial _i\varphi_0]$, which 
gives rise to the second-order Newtonian piece and is insensitive to any
non-Gaussianity in the initial conditions. The remaining pieces in Eqs.~(\ref{PSI}) and~(\ref{PHI}) correspond to contributions 
which tend to dominate on large scales with respect to those 
characterizing the Newtonian contribution, and whose origin is purely relativistic. 
In particular these are the pieces carrying the information on primordial 
non-Gaussianity. For a flat matter-dominated 
(Einstein-de Sitter) universe $g(\tau)=1$ and 
$B_1(\tau)=B_2(\tau)=0$, while $B_3(\tau) \rightarrow  (5/21) \tau^2$ and 
$B_4(\tau) \rightarrow  -\tau^2/14$, so that one recovers the expressions of~\cite{CMB2} (see also~\cite{BC}). 
In Ref.~\cite{tomita1} second-order cosmological 
perturbations have been computed in the 
$\Lambda \neq 0$ case from the synchronous to the Poisson gauge, 
thus extending the analysis of Ref.~\cite{MMB}, and the CMB temperature 
anisotropies induced by metric perturbations have been also considered by 
applying the expressions of Ref.~\cite{mm} (see also Ref.~\cite{Full}).
However, an important point to notice is that 
both Refs.~\cite{MMB,mm} and Ref.~\cite{tomita1} 
disregard any primordial non-linear contribution 
from inflation.\footnote{The results 
in Refs.~\cite{MMB,mm,tomita1} have initial conditions corresponding 
to our $a_{\rm NL}=0$.} 
     
The matter density contrast at second-order can now be calculated from the $(0-0)$ Einstein equation
\begin{eqnarray}
\label{002}
&& \fl 3 {\cal H}\psi^{(2)'}+3{\cal H}^2 \phi^{(2)}-\nabla^2 \psi^{(2)}+12{\cal H}\psi^{(1)}\psi^{(1)'}
-2 \nabla^2 \left( \psi^{(1)} \right)^2-12{\cal H}^2 \left( \phi^{(1)} \right)^2
-12{\cal H}\phi^{(1)}\psi^{(1)'} \nonumber \\
&& \fl -3\left( \psi^{(1)'}\right)^2
+\partial_i \psi^{(1)} \partial^i\psi^{(1)}-4\psi^{(1)}\nabla^2\psi^{(1)}=
-3{\cal H}^2 \Omega_m
\left( \frac{1}{2} \delta^{(2)} +v^{(1)2} \right)\, .
\end{eqnarray}

We thus arrive at the density contrast in the Poisson gauge   
\begin{eqnarray}
\label{delta2P}
\fl \delta^{(2)}&=& \frac{1}{\Omega_m}\Bigg[ (f-1)^2-\frac{2}{g^2} \frac{A'(\tau)}{{\cal H}}-\frac{2}{g^2} A(\tau)-1 \Bigg] 
\varphi^2-\frac{2}{\Omega_m g^2}\Bigg[ \frac{B'_2(\tau)}{{\cal H}}-\frac{4}{3}\frac{g'}{{\cal H}} 
g_{\rm in}+B_2(\tau)
\nonumber \\
\fl &-&\frac{4}{3}g g_{\rm in} +
\frac{4}{3}g^2 \left(e +\frac{3}{2} \right) \Bigg]\alpha({\bf x},\tau) -\frac{2}{\Omega_m g^2}\frac{B'_3(\tau)}{{\cal H}}
\nabla^{-2} \partial_i\partial^j(\partial^i \varphi \partial_j \varphi ) - \frac{2}{\Omega_m g^2} \frac{B'_4(\tau)}{{\cal H}}
\partial_i \varphi \partial^i \varphi\nonumber \\
\fl &+& \frac{2}{3 \Omega_m g^2}\Bigg[ B_1(\tau)-2 g g_{\rm in} -\frac{10}{3} (a_{\rm NL}-1)g g_{\rm in}+2 g^2 \Bigg] 
\frac{\nabla^2 \varphi^2}{{\cal H}^2} + \frac{2}{3 \Omega_m g^2} \Bigg[ 
B_2(\tau)- \frac{4}{3} g g_{\rm in}
\nonumber \\
\fl &-& \frac{4}{3}\frac{f^2g^2}{\Omega_m}-g^2-3 {\cal H}^2 B_4(\tau) \Bigg] 
\frac{\partial_i \varphi \partial^i \varphi}{{\cal H}^2} 
+ \frac{2}{3 \Omega_m g^2} \Bigg[ 
-3 B_2(\tau)+4g g_{\rm in} -3 {\cal H}^2 B_3(\tau) \Bigg] 
\nonumber \\
\fl &\times& \frac{1}{{\cal H}^2} 
\nabla^{-2} \partial_i\partial^j(\partial^i \varphi \partial_j \varphi )+\frac{8}{3 \Omega_m} 
\frac{\varphi \nabla^2 \varphi}{{\cal H}^2}
+ \frac{2}{3 \Omega_m g^2} \frac{B_3(\tau)}{{\cal H}^2} \partial_i\partial^j(\partial^i \varphi \partial_j \varphi )+
\frac{2}{3 \Omega_m g^2} \nonumber \\
\fl &\times& 
\frac{B_4(\tau)}{{\cal H}^2} \nabla^2(\partial^i \varphi \partial_i \varphi) \, ,
\end{eqnarray}
where $A(\tau)\equiv [B_1(\tau)-2g g_{\rm in}-(10/3) (a_{\rm NL}-1) g g_{\rm in}]$ and $\alpha({\bf x},\tau)$ has the same 
expressions as $\alpha_0({\bf x})$ introduced in Eq.~(\ref{alpha0}) with $\varphi$ in place of $\varphi_0$. Notice that we have 
explicitly verified that Eq.~(\ref{delta2P}) in the limit of an Einstein-de Sitter universe recovers the expression for the matter density contrast obtained in Ref.~\cite{noiLSS}.  

Eq.~(\ref{delta2P}) is the main result of this section. It shows how the primordial NG, which is initially generated 
on large scales, is transferred to the density contrast on subhorizon scales. The expression for 
the density contrast is made of three contributions: a second-order
Newtonian piece (the last two terms in Eq.~(\ref{delta2P}), proportional to $\tau^4$ in an Einstein-de Sitter universe) 
which is insensitive to the non-linearities in the initial conditions; a Post-Newtonian (PN)
piece (related to two gradient terms of the gravitational potential) which carries the most relevant
information on primordial NG; the super-horizon terms (corresponding to the first two lines of Eq.~(\ref{delta2P})).  
Our findings show in a clear way that the information on the
primordial NG set on super-Hubble scales flows into the PN terms, leaving
an observable imprint in the LSS. Another interesting result which shows up in Eq.(\ref{delta2P}) is the presence 
in the Post-Newtonian term 
of contributions different from the primordial NG which are due to weakly non-linear corrections which switch on 
when the modes cross inside the Hubble 
radius and which can constitute an interesting additional source of non-Gaussianity since they can probe large-scale 
GR corrections.  

\subsection{The non-linearity parameter $f_{\rm NL}$} 

We can rewrite the matter density contrast in Fourier space in 
terms of the linear density contrast as in Eq.~(\ref{defkerneldensity}). We find   
\begin{eqnarray}
\label{KN}
{\cal K}_{\delta}^N({\bf k}_1,{\bf k}_2;\tau)  & = &  
\frac{3}{4} \frac{\Omega_m}{g^2}\, {\cal H}^2 \left[ B_3(\tau) \frac{({\bf k}\cdot {\bf k}_1) ({\bf k}\cdot {\bf k}_2)}{k_1^2 k_2^2} + B_4(\tau) 
\frac{k^2({\bf k}_1\cdot {\bf k}_2)}{k_1^2 k_2^2} \right]\, ,
\end{eqnarray}
and for the non-linearity parameter 
\begin{eqnarray}
\label{fnldd}
\fl f^{P}_{\rm NL}({\bf k}_1,{\bf k}_2;\tau) &=& \Bigg[ \frac{5}{3} (a_{\rm NL}-1)+1-\frac{g}{g_{\rm in}}-\frac{1}{2} \frac{B_1(\tau)}{gg_{\rm in}} \Bigg]
-\frac{\left(k_1^2+k_2^2\right)}{k^2} \frac{g}{g_{\rm in}}\\
\fl &+& \frac{({\bf k}_1\cdot {\bf k}_2)}{k^2} \Bigg[ 
\frac{2}{3}\, e(\tau)\, \frac{g}{g_{\rm in}}+\frac{1}{2} \frac{g}{g_{\rm in}}+\frac{2}{3}-\frac{1}{2} \frac{B_2(\tau)}{g g_{\rm in}}+\frac{3}{2} {\cal H}^2 \frac{B_4(\tau)}{g g_{\rm in}} \Bigg]
\nonumber \\
\fl &+& \frac{({\bf k}\cdot {\bf k}_1) ({\bf k}\cdot {\bf k}_2)}{k^4} \Bigg[ \frac{3}{2} {\cal H}^2 \frac{B_3(\tau)}{g g_{\rm in}}-2+\frac{3}{2} \frac{B_2(\tau)}{g g_{\rm in}} \Bigg]
\nonumber \\
\fl &-& \frac{({\bf k}\cdot {\bf k}_1) ({\bf k}\cdot {\bf k}_2)}{k^2} \frac{k_1^2+k_2^2}{k_1^2k_2^2}\,  \frac{3}{2} {\cal H}^2 f(\tau) \frac{B_3(\tau)}{gg_{\rm in}}- 
({\bf k}_1\cdot {\bf k}_2) \frac{k_1^2+k_2^2}{k_1^2k_2^2}\, \frac{3}{2} {\cal H}^2 f(\tau) \frac{B_4(\tau)}{gg_{\rm in}} \nonumber \, .
\end{eqnarray} 
The non-linearity parameter $f_{\rm NL}$ is defined via Eq.~(\ref{P}) and in this
way it generalizes the standard definition of Ref.~\cite{verde} inferred from the Newtonian gravitational potential. 
In order to obtain the expressions in Eqs.~(\ref{KN}) and~(\ref{fnldd}) we have performed an 
expansion in $({\cal H}/ k_{1,2})\ll 1$ up to terms $({\cal H}/ k_i)^{2}$ 
starting from Eq.~(\ref{delta2P}). In Eq.~(\ref{fnldd}) the primordial NG is clearly evident in the piece proportional 
to $(a_{\rm NL}-1)$. The remaining terms are due to the horizon scale Post-Newtonian corrections we commented about in Eq.~(\ref{delta2P}). The non-linearity (NG) induced 
by these terms show a specific and non-trivial shape dependence that can help in detecting them. Of course their relative importance increases with the scale, and in fact it has been 
shown in Ref.~\cite{VMb} that, through the large-scale halo bias techniques investigated in Refs.~\cite{Dalal,MatVerde}, these GR corrections are potentially detectable. 

It is also worth noticing that some of the terms entering in the non-linearity parameter vanish in the limit of a  vanishing cosmological constant (e.g.
$B_1(\tau)$ and $B_2(\tau)$ go to zero in this limit).
   
A final comment on Eq.~(\ref{defkerneldensity}). In this expression, 
when the various modes are well inside the horizon, one can take the usual configuration for the linear density perturbations with 
$\delta^{(1)}_{\bf k}(\tau)\propto (2k^2 T(k)/3 \Omega_{m0} H^2_0) D_{+}(a)$, where $T(k)$ is the usual linear matter transfer function. In this way one partially 
accounts for the effects of the transition from a radiation- to a matter-dominated epoch. In fact a full computation would require to study up to second-order the 
evolution of the density perturbations also during the radiation dominated era. In this way one can recover a full matter 
transfer function up to second-order. Details about such a computation can be found in Refs.~\cite{CMB2,pitrou,Fitz}.  
For example, Ref.~\cite{Fitz} shows that the corrections from the full matter 
transfer function, when accounting for a matching at second-order to the radiation epoch, gives a relative correction with respect to the Newtonian 
kernel that is of order $a_{\rm eq}\sim 10^{-4}$ and that such a correction is equivalent to the effect of a primordial NG of 
$f_{\rm NL}\sim 4$. That these small-scale corrections are tiny is easy to understand. By looking at Eq.~(\ref{002}) one realizes that accounting for the 
radiation epoch at second-order the matter density perturbations from, say approximately the matter-radiation equality epoch onwards, will 
get a correction which scales like $\delta^{(2)}|_{\rm equiv} \, \tau^2$, which rescales the matching initial conditions to the radiation epoch. However we expect such term to be negligible 
w.r.t. the Newtonian part  which scales like $(\delta^{(1)}_{\bf k}(\tau))^2 \propto \tau^4$, and also w.r.t. a sizable primordial NG.

\section{Dark matter density perturbations at second-order: comoving-synchronous gauge}

Let us now see how the second-order matter perturbations and gravitational potentials 
are obtained in the comoving-synchronous gauge. 
 
We make use of the formalism developed in Refs.~\cite{mps1,mps2,MT,MMB} which the reader is referred to
for more details.   
The synchronous, time-orthogonal gauge is defined by setting $g_{00}=-a^2(\tau)$ and $g_{0i}=0$, so that the line-element 
takes the form 
\begin{equation}
ds^2=a^2(\tau)[-d\tau^2+\gamma_{ij}({\bf x},\tau) dx^i dx^j]\, .
\end{equation} 
For our fluid containing irrotational, pressure-less matter plus $\Lambda$, this also 
implies that the fluid four-velocity field is given by   $u^{\mu}=(1/a,0,0,0)$, so that $x$ represent 
comoving ``Lagrangian" coordinates for the fluid element (indeed, the possibility of making the synchronous, time-orthogonal 
gauge choice and comoving gauge choice simultaneously is a peculiarity of fluids with vanishing spatial pressure gradients, i.e. 
vanishing acceleration, which holds at any time, i.e. also beyond the linear regime). 
A very efficient way to write down  Einstein and continuity equations is 
to introduce the peculiar velocity-gradient tensor~\cite{MT} 
\begin{equation}
\vartheta^{i}_j\equiv 
u^{i}_{~;j}-\frac{a'}{a}\delta^{i}_{j}=\frac{1}{2}\gamma^{ik}
\gamma'_{kj} \;, 
\end{equation}
where we have subtracted the isotropic Hubble flow. Here semicolons  
denote covariant differentiation. 
From the continuity equation $T^{\mu\nu}_{~~~;\nu}=0$, we infer  
the  exact solution for the density contrast $\delta=\delta \rho/\rho$~\cite{MT,MMB}
\label{delta}   
\begin{equation}
\label{delta}
\delta({\bf x, \tau})=(1+\delta_0({\bf x}))[\gamma({\bf x},\tau)
/\gamma_0({\bf x})]^{-1/2}-1\, ,
\end{equation}
where $\gamma={\rm det} \gamma_{ij}$. The subscript ``$0$'' denotes the 
value of quantities evaluated at the present time.  
 
From Eq.~(\ref{delta}) it is evident that in the comoving-synchronous
gauge the only independent 
degree of freedom is the spatial metric tensor $\gamma_{ij}$.
The energy constraint reads
\begin{equation}
\label{energycon}
\vartheta^2-\vartheta^i_{~j}\vartheta^j_{~i}+4{\cal H} \vartheta
+ {\mathcal R} = 6 {\cal H}^2 \Omega_m \delta \, ,
\end{equation}  
where ${\mathcal R}^i_{~j}$ is the Ricci tensor associated with the spatial 
metric $\gamma_{ij}$ with scalar curvature ${\mathcal R}={\mathcal R}^i_{~i}$.
The momentum constraint reads
$\vartheta^i_{~j|i}=\vartheta_{,j}$, where bars stand for covariant 
differentiation in the three-space with metric
$\gamma_{ij}$. Finally, one can use the Raychaudhuri equation 
\begin{equation}
\label{Ray}
\vartheta'+{\cal H} \vartheta+\vartheta^i_{~j}\vartheta^j_{~i}+
\frac{3}{2}{\cal H}^2 \Omega_m \delta=0\, ,
\end{equation} 
which is obtained from the energy constraint and the trace of the 
evolution equation
\begin{equation}
\label{EVO}
\vartheta^{i'}_{~j}+2 {\cal H} \vartheta^i_{~j} 
+\vartheta\vartheta^{i}_{~j} + \frac{1}{4} \left(\vartheta^k_{~l} \vartheta^l_{~k}-\vartheta^2 \right)
\delta^i_{~j} + {\mathcal R}^i_{~j}
- \frac{1}{4} {\mathcal R} \delta^i_{~j} = 0 \, .
 \end{equation} 
Notice that these equations are exact 
and describe the fully non-linear evolution of cosmological perturbations 
(up to the time of caustic formation). 
In order to show how the primordial non-Gaussianities appear in 
in the matter density contrast,  we then perform a 
perturbative expansion up to second order in the fluctuations of the 
metric. 

The spatial metric tensor can  be expanded as
\begin{equation}
\gamma_{ij}=(1-2\psi^{(1)}-\psi^{(2)})\, \delta_{ij} + \chi^{(1)}_{ij}+\frac{1}{2} 
\chi^{(2)}_{ij}\, , 
\end{equation}
where $\chi^{(1)}_{ij}$ and $\chi^{(2)}_{ij}$ are 
traceless tensors and include scalar, vector and tensor (gravitational waves) 
perturbations. As usual we split 
the density contrast into a linear and a second-order part as 
$\delta({\bf x},\tau)=\delta^{(1)}({\bf x},\tau)+\frac{1}{2}
\delta^{(2)}({\bf x},\tau)$. At linear order the growing-mode solutions 
in the comoving-synchronous gauge are given by
\begin{eqnarray}
\psi^{(1)}({\bf x},\tau) & = & \frac{5}{3}\varphi_{\rm in}({\bf x}) + \frac{2}{9 {\cal H}^2(\tau) \Omega_m(\tau)} \nabla^2 \varphi({\bf x},\tau)\, ,  \nonumber \\
\chi^{(1)}_{ij} ({\bf x},\tau) & = & D_{ij} \chi^{(1)} ({\bf x},\tau), \ \ \  \ \ \ \ \  \chi^{(1)}({\bf x},\tau) 
= -  \frac{4}{3 {\cal H}^2(\tau) \Omega_m(\tau)} \varphi({\bf x},\tau)\, , 
\end{eqnarray}
where $\varphi({\bf x},\tau)$ is the growing-mode scalar potential defined in Eq.~(\ref{relphiphi_0}). 
The linear density contrast $\delta^{(1)}$ in this gauge is related to $\varphi$ via the usual Poisson equation, namely  
\begin{equation}
\nabla^2 \varphi({\bf x},\tau) = \frac{3}{2} {\cal H}^2(\tau) \Omega_m(\tau) \delta^{(1)}({\bf x},\tau)  \, . 
\end{equation} 
 
In writing $\chi^{(1)}_{ij}$ we have eliminated the residual gauge ambiguity of the synchronous gauge
as in Ref.~\cite{MMB}  \footnote{More in general, at any order $n$ in perturbation theory the scalar potentials 
$\psi^{(n)}$ and $\chi^{(n)}$ can be shifted by arbitrary constant amounts  $\delta\psi^{(n)}_0$ and $\delta\chi^{(n)}_0$, only provided 
$\delta\psi^{(n)}_0 + (1/6) \nabla^2 \delta\chi^{(n)}_0 = 0$.}. 
We have also assumed that linear vector modes are absent, since they are not produced in standard 
mechanisms for the generation of cosmological perturbations (as inflation). 
We have also neglected linear tensor modes, since they play a negligible
role in LSS formation. 

By perturbing Eq.~(\ref{delta}) up to second order, 
we get 
\begin{eqnarray}
\label{delta2s}
\delta^{(2)} & = & \delta^{(2)}_0 + 3 \left(\psi^{(2)}-\psi^{(2)}_0\right) 
+\frac{1}{2}\left(D_+^2 -1\right) \bigg[ \frac{1}{2} \left(\nabla^2 \chi^{(1)}_0\right)^2 + 
\nonumber \\
&+& \partial^i \partial^j \chi^{(1)}_0 \partial_i \partial_j \chi^{(1)}_0 \bigg]
- \left(D_+ -1\right) \left( 2 \psi^{(1)}_0 + \frac{1}{3} \nabla^2 \chi^{(1)}_0 \right) \nabla^2 \chi^{(1)}_0 \, .
\end{eqnarray}

To compute the metric perturbation $\psi^{(2)}$, we can use the evolution equation, Eq.~ (\ref{EVO}). To this aim
it proves convenient to write 
\begin{eqnarray}
\label{psi2xi}
\psi^{(2)} & = & \psi^{(2)}_0 - \frac{1}{3} \delta^{(2)}_0 + \frac{1}{6} \partial^i \partial^j \chi^{(1)}_0 \partial_i \partial_j \chi^{(1)}_0 - \nonumber \\
& - & \frac{1}{3} \left( 2 \psi^{(1)}_0 + \frac{1}{3} \nabla^2 \chi^{(1)}_0 \right) \nabla^2 \chi^{(1)}_0 + \frac{1}{12} \left(\nabla^2 \chi^{(1)}_0\right)^2 + \xi \;,
\end{eqnarray}
where we have introduced the variable $\xi$ which is determined by the equation
\begin{eqnarray}
&& \xi^{\prime\prime} + {\cal H} \xi^\prime - \frac{3}{2} {\cal H}^2 \Omega_m \xi = \nonumber \\
&& = \frac{1}{4} {\cal H}^2 \Omega_m D_+^2 \left[ \frac{1}{2} \left( \nabla^2 \chi^{(1)}_0 \right)^2 - 
\left( 1 +  \frac{2}{3 e(\Omega_m)} \right) \partial^i \partial^j \chi^{(1)}_0 \partial_i \partial_j \chi^{(1)}_0 \right] \;.
\end{eqnarray}

An approximate solution of this equation can be obtained by making use of the usual approximation 
(see e.g. Ref.~\cite{colnus,Bernard}) $e(\Omega_m) \approx 1$, which holds true for reasonable values of $\Omega_m$.   
Under these circumstances we can write
\begin{equation}
\xi(\tau) \approx A D_+(\tau) + \frac{D_+^2(\tau)}{14} \left[  \frac{1}{2} \left( \nabla^2 \chi^{(1)}_0 \right)^2 - 
\frac{5}{3} \partial^i \partial^j \chi^{(1)}_0 \partial_i \partial_j \chi^{(1)}_0 \right] \;,
\end{equation}
where $A$ is an integration constant which can be determined by the energy constraint at the initial 
time. Notice that it is precisely the sub-leading, Post-Newtonian term proportional to the linear growing mode $D_+$ 
which brings all the relevant information about primordial and GR-induced non-Gaussianity. 
It is also important to stress that the time
dependence of this term, which comes from the homogeneous solution of the above equation is {\it exact}, while 
the above approximation only affects the fastest growing Newtonian terms, i.e. those proportional to 
$D_+^2$. 
Using this procedure and providing the initial data (formally at $\tau \to 0$) in terms of the gauge-invariant curvature perturbation, which in this gauge 
reads $\zeta^{(1)}_{\rm in} = - \psi^{(1)}_{\rm in} - (1/6) \nabla^2 \chi^{(1)}_{\rm in} = - 5 \varphi_{\rm in}/3 $ and $\zeta^{(2)}_{\rm in} 
= - \psi^{(2)}_{\rm in} - (1/6) \nabla^2 \chi^{(2)}_{\rm in} = 
50 a_{\rm nl} \varphi^2_{\rm in}/9$, we finally obtain
\begin{eqnarray}
\fl \delta^{(2)}(\tau) & = & \frac{100}{9{\cal H}_0^2} \left[f(\Omega_{0m}) + \frac{3}{2} \Omega_{0m}\right]^{-1} \bigg\{ D_+(\tau)\left[\left(\frac{3}{4} 
- a_{\rm NL}\right)\left(\nabla \varphi_{\rm in}\right)^2 
+ (2 - a_{\rm NL} ) \varphi_{\rm in} \nabla^2 \varphi_{\rm in} \right]
\nonumber  \\
\fl & + & \frac{D_+^2(\tau)}{14 {\cal H}^2_0} \left[f(\Omega_{0m}) + \frac{3}{2} \Omega_{0m}\right]^{-1} \left[ 5 \left( \nabla^2 \varphi_{\rm in} \right)^2 +
2 \partial^i \partial^j \varphi_{\rm in}  \partial_i \partial_j \varphi_{\rm in} \right] \bigg\}\;,
\end{eqnarray}
where we made use of the following property 
\begin{equation}
{\cal H} D_+^\prime + \frac{3}{2} {\cal H}^2 \Omega_m D_+ = {\rm const.} \;,
\end{equation} 
whose validity can be easily proven on the basis of Eq.~(\ref{D+}) and of the Friedmann equation
${\cal H}^\prime - {\cal H}^2 + (3/2)  {\cal H}^2 \Omega_m =0$. One may notice that even the Newtonian part (i.e. the one proportional to
$D_+^2$) of the second-order matter perturbations 
differs from the Poisson gauge expression. This is a well-known feature that is to be ascribed to the different meaning 
of the mass density when moving from Eulerian to Lagrangian coordinates, which only appears at second and higher orders 
(see, e.g. Ref.~\cite{bouchet}). 

As we did in the Poisson-gauge case, we can re-express this result in terms of a suitable 
potential $\Phi$ in order to introduce a non-linearity parameter $f_{\rm NL}$ with the usual meaning. 
We obtain (see also Ref.~\cite{VMb})
\begin{equation}
f^{C}_{\rm NL}({\bf k}_1,{\bf k}_2)  =  \frac{5}{3} \left[ (a_{\rm NL}-1) - 1 + \frac{5}{2} \frac{{\bf k}_1 \cdot {\bf k}_2}{k^2} \right]
\end{equation} 
Notice that the non-linearity parameter is different w.r.t. the one in Eq.~(\ref{fnldd}), not only because they have been computed in two different gauges, but also because 
in the comoving synchronous gauge the Post-Newtonian term giving rise to $f^{S}_{\rm NL}$ can be easily written in an exact form where the growing mode $D_+(\tau)$ is 
factored out, while this is not the case for the Poisson gauge. 
It is also important to stress here that the expression for $f_{\rm NL}$ obtained in this gauge is the one to be used to 
evaluate the effect of NG on the Lagrangian bias of dark matter halos, as recently stressed in Ref.~\cite{wanslo}. 
In Ref.~\cite{VMb} it was shown that tiny effects in the above equation which come purely from the General Relativistic evolution, i.e. 
the term which survives in the limit $a_{\rm NL} =1$, are potentially detectable for some planned LSS surveys. 

\section{Concluding remarks}

In this paper we have investigated the effect of primordial and GR-induced non-Gaussianities on the 
second-order matter density perturbation in a $\Lambda$CDM cosmology. 
The calculation has been performed in two popular gauges, the Poisson gauge and 
the comoving time-orthogonal one, which are useful for comparison with observations, depending on the particular
quantity under study. For instance, in evaluating the effect of NG on the mass function and Lagrangian bias of dark matter halos, 
the comoving gauge expression is more appropriate, while the Poisson gauge formulae are more suitable for
gravitational lensing studies.  
The strongest present limits on $f_{\rm NL}$ come from the analysis of the angular bispectrum of WMAP 
temperature anisotropy data. Indeed, Komatsu et al. \cite{k10}, analyzing the 7-years WMAP data obtain  
the $95\%$ limits $- 10 < f_{\rm NL}^{\rm local} < 74$,  and $- 214 < f_{\rm NL}^{\rm equilateral} < 266$. 
The analysis of Planck data both in temperature and E-mode polarization is expected to improve the 
accuracy by almost an order of magnitude. 
A complementary and very powerful information on the amplitude and shape of primordial NG will come from the study the galaxy clustering 
(e.g. Ref.~\cite{verde2010}) and other LSS datasets, such as weak gravitational lensing (e.g. Refs~\cite{lesg,fede}) and 
redshifted 21cm background anisotropy (e.g. Refs.~\cite{pille1,coor2}).
Primordial NG in LSS data can be searched for by various techniques: abundance of massive and/or 
high-redshift objects \cite{MVJ,vjkm,LoVerde,grossi,jimen,MR1,MR2}, abundance of voids \cite{jkv}, 
higher-order statistics such as bispectrum and trispectrum (see, e.g. Ref.~\cite{revliguori}), large-scale clustering of halos, 
thought as rare high peaks of the dark matter distribution \cite{MLB,Dalal,MatVerde,desjacques,pille2,grossi}, 
and their cross-correlation with the CMB via the Integrated Sachs-Wolfe effect \cite{Afshordi}, small-scale NG corrections 
to the matter power-spectrum \cite{pille2,juan}. 
For instance, Slosar et al. \cite{slosar}, exploiting the scale-dependence of the NG correction to the halo linear 
bias $\Delta b_{\rm NG} \propto f_{\rm NL}^{\rm local} k^{-2}$, obtained the 
$95\%$ confidence range $- 29 <   f_{\rm NL}^{\rm local} < 70$, using the two-point function of a combination of datasets. 
The prospects for sensibly narrowing  these limits with the advent of galaxy surveys sampling regions comparable 
to the Hubble volume are analyzed in Ref.~\cite{carbone}. Indeed, as shown in Ref.~\cite{VMb}, there are very promising 
prospects to observe NG signatures down to the limits of the order unity GR corrections discussed in this paper. This largely motivates 
theoretical efforts to  obtain accurate predictions of these effects. 

\section*{Acknowledgments}
A.R. acknowledges partial support by the EU Marie Curie Network UniverseNet (HPRNCT2006035863).
This research has been partially supported by the ASI Contract No.  I/016/07/0 COFIS, the ASI/INAF
Agreement I/072/09/0 for the Planck LFI Activity of Phase E2. 


\end{document}